\documentclass[a4paper,twocolumn]{article}
\usepackage[T1]{fontenc}
\usepackage{graphics}

\makeatletter

\providecommand{\LyX}{L\kern-.1667em\lower.25em\hbox{Y}\kern-.125emX\@}
\newcommand{\noun}[1]{\textsc{#1}}
\let\SF@@footnote\footnote
\def\footnote{\ifx\protect\@typeset@protect
    \expandafter\SF@@footnote
  \else
    \expandafter\SF@gobble@opt
  \fi
}
\expandafter\def\csname SF@gobble@opt \endcsname{\@ifnextchar[
  \SF@gobble@twobracket
  \@gobble
}
\edef\SF@gobble@opt{\noexpand\protect
  \expandafter\noexpand\csname SF@gobble@opt \endcsname}
\def\SF@gobble@twobracket[#1]#2{}

\usepackage{vmargin}

\makeatother

\begin{document}

\title{Strangeness Suppression in Proton-Proton Collisions}

\author{Hans-Joachim Drescher\protect\( ^{\$}\protect \), Jörg Aichelin\protect\( ^{\dagger }\protect \),
Klaus Werner\protect\( ^{\dagger }\protect \)\\
\protect\( ^{\$}\protect \)New York University, 4 Washington Place, New York
NY 10003\\
\protect\( ^{\dagger }\protect \)SUBATECH, La Chantrerie 4, rue Alfred Kastler
BP 20722 - 44307 Nantes-cedex 3}

\maketitle
\begin{abstract}
We analyse strangeness production in proton-proton (pp) collisions at SPS and
RHIC energies, using the recently advanced NeXus approach. After having verified
that the model reproduces well the existing data, we interpret the results:
strangeness is suppressed in proton-proton collisions at SPS energy as compared
to electron-positron (\( \mathrm{e}^{+}\mathrm{e}^{-} \)) annihilation due
to the limited masses of the strings produced in the reaction, whereas high
energy pp and \( \mathrm{e}^{+}\mathrm{e}^{-} \) collisions agree quantitatively.
Thus strangeness suppression at SPS energies is a consequence of the limited
phase-space available in string fragmentation. 
\end{abstract}

\section{Introduction}

Strangeness enhancement in ultra-relativistic nucleus-nucleus collisions has
been proposed as a signal for the formation of a quark-gluon plasma\cite{rafelski}.
At the SPS it has been observed for example by the WA97 \cite{WA97} collaboration,
and the experiments of RHIC will certainly analyse strangeness is detail. When
one talks about strangeness enhancement one has first to specify the point of
reference which are usually proton-proton reactions. How meaningful is this
point of reference? In order to answer this question, we analyse the strangeness
production of proton-proton collisions at different energies as compared to
\( \mathrm{e}^{+}\mathrm{e}^{-} \)annihilation which is rather energy independent.
This choice is justified by the fact that particle production seems to be universal
in all kinds of elementary high energy reactions.

\section{The \noun{neXus} Model and the Role of String Fragmentation}

Before we start with an analysis of the physics of multiplicities of different
hadrons we explain the basic features of our approach (\noun{neXus}) which
describes simultaneously high energy electron-positron annihilation and hadron-hadron
scattering. The details may be found in reference \cite{nexus}. 

The common feature between hadron-hadron collisions and electron-positron annihilation
is the creation of strings which finally produce observable hadrons. In the
former case the exchange of a Pomeron leads to the formation of two strings,
in the latter a string is spanned between the quark-antiquark created by the
decay of a virtual photon or a Z-boson. At low energies the string just consists
of two partons at the end-points, at higher energies perturbative gluons appear
in initial or final state radiation which are mapped onto the string as the
so-called kinks.

\begin{figure}
{\par\centering \resizebox*{1\columnwidth}{!}{\includegraphics{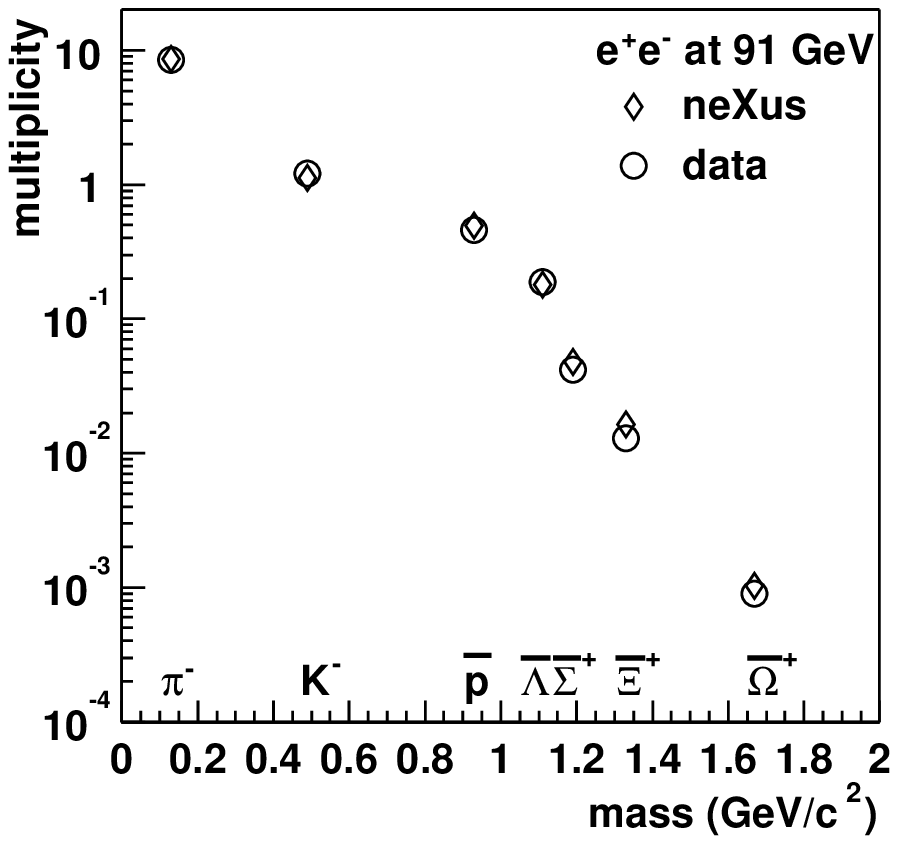}} \par}

\caption{\label{fig:1a}Results for \protect\( \mathrm{e}^{+}\mathrm{e}^{-}\protect \)
annihilation at 91.2 GeV compared with data from the Opal collaboration \cite{opal}.}

{\par\centering \resizebox*{1\columnwidth}{!}{\includegraphics{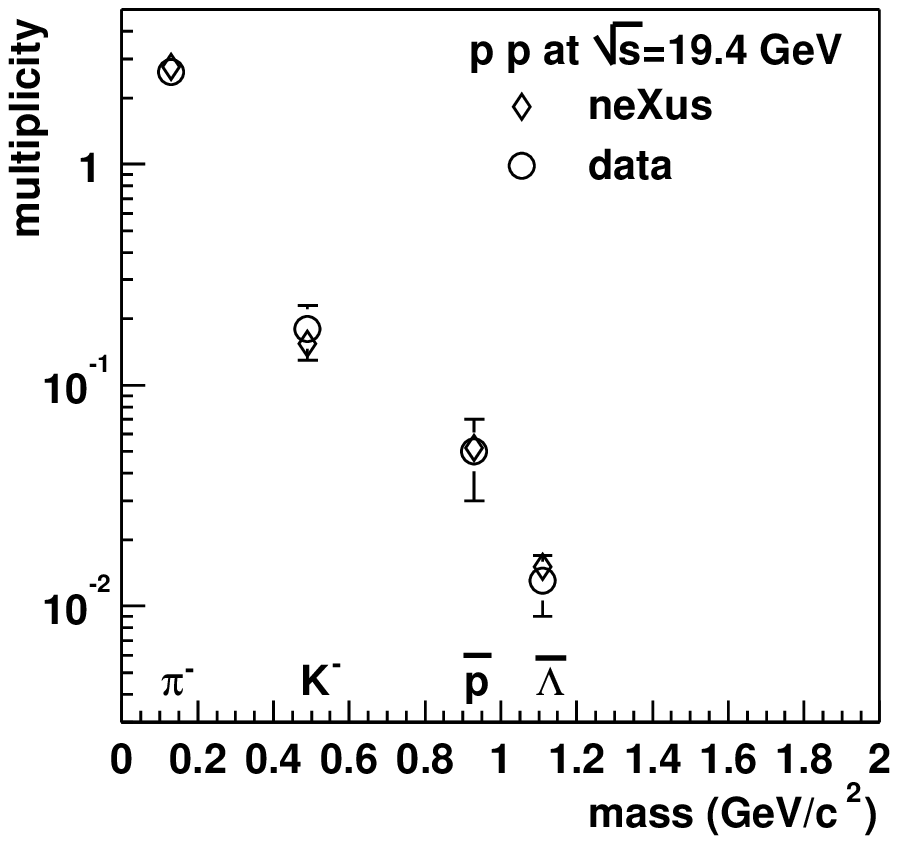}} \par}

{\par\centering \resizebox*{1\columnwidth}{!}{\includegraphics{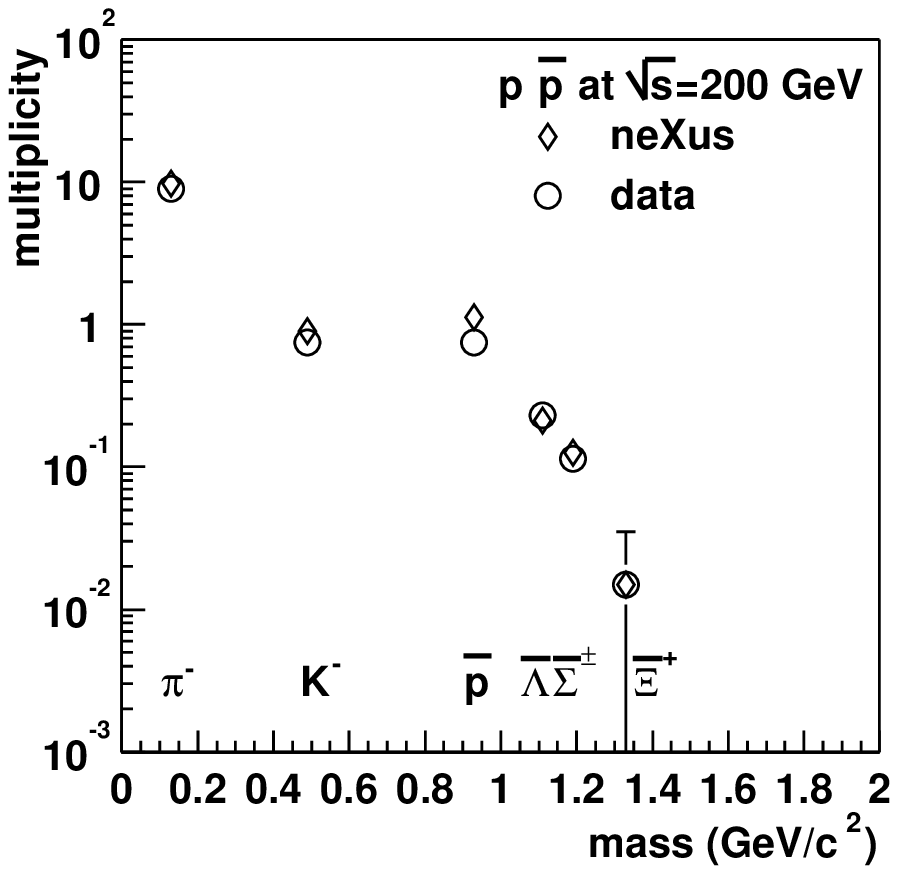}} \par}

\caption{\label{fig:1bc}Comparison of the model with data for proton proton collisions
at \protect\( \sqrt{s}=19.4\protect \) GeV\cite{pp19} and anti-proton-proton
collisions at \protect\( \sqrt{s}=200\protect \) GeV \cite{pp200}.}
\end{figure}
Once a string is created it evolves according to the Nambu-Goto Lagrangian for
a classical relativistic string. In order to produce hadrons we use the area-law
of Artru-Menessier. Here, the probability of the string to break is proportional
to the invariant area swept over in Minkowski space. The breaking is then determined
by one parameter, the break probability. If it is small, the string breaks at
later times, producing less but heavier fragments and vice-verse. Flavor production
is governed by two additional parameters, the probability to create a strange
quark-antiquark pair (otherwise up or down pairs are created in equal fractions)
and the probability to create a diquark-antidiquark pair. The former therefore
governs strangeness production, the latter baryon production and the combination
of both rules the creation of hyperons. The decay of strings can be seen as
a longitudinal (one dimensional) microscopic phase space decay, therefore is
is more difficult to produce heavier particles than lighter ones. 

In figure \ref{fig:1a} we show particle yields of \( \mathrm{e}^{+}\mathrm{e}^{-} \)
annihilation from our model compared with data from the Opal collaboration \cite{opal}.
The two parameters for strangeness and diquarks have been adjusted to fit these
data, and the model is capable to describe a multitude of data. One can convince
oneself in reference \cite{nexus} that also event-shapes and differential spectra
are reproduced nicely. The same model applied to hadron-hadron collisions gives
the results shown in figure \ref{fig:1bc}. Here we compare two energies which
are close to the ones we are going to use for our analysis. Furthermore we consider
only negatives or anti-baryons as produced particles, results for the other
particles agree in a similar way with data. We can conclude that for \( e^{+}e^{-} \)annihilation
as well as for pp and anti-proton-proton (\( \mathrm{p}\bar{\mathrm{p}} \))
collisions NeXus agrees with the experimentally observed particle yields.

\section{Interpretation of Results}

\begin{figure}
{\par\centering \resizebox*{1\columnwidth}{!}{\includegraphics{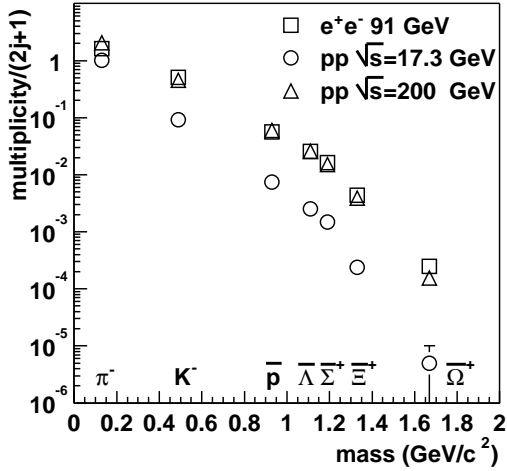}} \par}

\caption{\label{fig:2}Particle yields of different reactions calculated with \noun{neXus}.}
\end{figure}

We are now going to interpret the above-mentioned results based on \noun{neXu}s
calculations. In figure \ref{fig:2} we show multiplicities of particles produced
in pp collisions at 17.3 GeV (SPS) and at 200 GeV (RHIC) as compared with \( \mathrm{e}^{+}\mathrm{e}^{-} \)
at 91.2 GeV. To account for spin-degeneracy we divide the obtained multiplicity
by the factor \( 2j+1 \). First of all one sees that the particle yields fall
roughly exponentially with the particle mass. This is a simple phase-space effect:
heavier particles are more difficult to produce. Striking is the unexpected
similarity of pp at 200 GeV with \( \mathrm{e}^{+}\mathrm{e}^{-} \) at 91.2
GeV. As described above, the formation of strings is quite different in pp as
compared to \( \mathrm{e}^{+}\mathrm{e}^{-} \) reactions. The spectra obtained
for 17.3 GeV is considerably steeper. We see as well very little difference
between strange and non strange hadrons, all fall on a common curve.

\begin{figure}
{\par\centering \resizebox*{1\columnwidth}{!}{\includegraphics{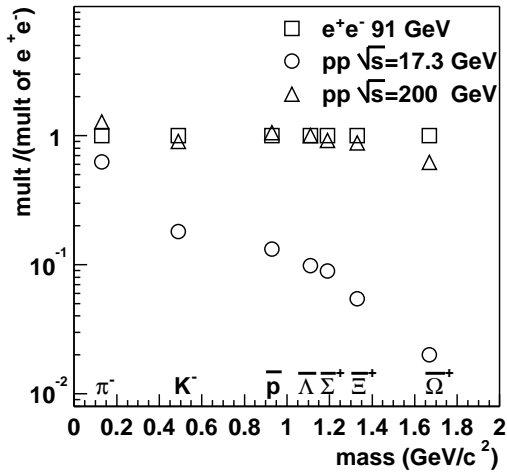}} \par}

\caption{\label{fig:3}The multiplicities of particles normalized to \protect\( \mathrm{e}^{+}\mathrm{e}^{-}\protect \)
at 91.2 GeV. Proton-proton collisions at 200 GeV are very similar to \protect\( \mathrm{e}^{+}\mathrm{e}^{-}\protect \),
at 17.3 GeV a suppression of heavier particles is noticeable. }
\end{figure}

This effect can be seen clearer in figure \ref{fig:3}, where the multiplicities
are plotted normalized to the ones of \( \mathrm{e}^{+}\mathrm{e}^{-} \). The
ratio for pp interactions with respect to \( \mathrm{e}^{+}\mathrm{e}^{-} \)
at RHIC energies is close to one. Only the heaviest particle - the Omega - is
slightly suppressed in pp. At SPS energies the yields for pp collisions show
a completely different behavior: the ratio with respect to \( \mathrm{e}^{+}\mathrm{e}^{-} \)
falls off strongly as a function of the mass.

Where does this effect come from? No new physics enters between the two energies,
with exception of the minijets which are more abundant at higher energies. But
this influences only differential spectra like that of transverse momenta and
not the relative abundance of particles. 

\begin{figure}
\resizebox*{1\columnwidth}{!}{\includegraphics{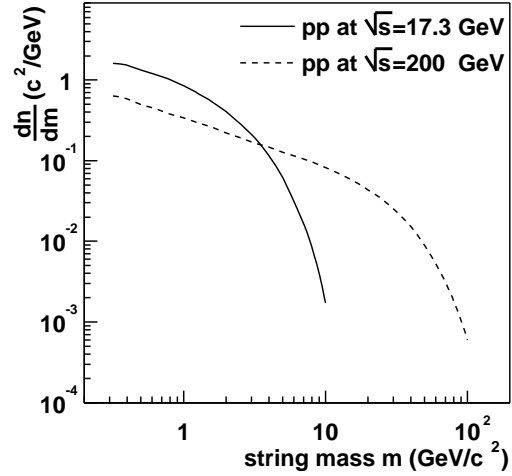}}

\caption{\label{fig:4}The distribution of string masses for two reactions pp at 17.3
GeV and 200 GeV. }
\end{figure}
The answer becomes quite clear when we look at the masses of the strings which
finally produce the particles. Figure \ref{fig:4} shows the distribution \( \frac{dn}{dm} \)
of string masses produced at the two different energies. We leave out the case
of \( \mathrm{e}^{+}\mathrm{e}^{-} \) since here we have in most cases one
string of mass 91.2 GeV. Only if a quark-anti-quark pair is produced during
the final state radiation, we end up with more than one string. This process
is however much less important than gluon radiation. In pp interactions most
of the strings have still low masses, which is a direct consequence of parton
distribution functions peaking at low x. But the evolution of the tails is quite
different. Whereas at 17.3 GeV the distribution is steeply falling with almost
no strings at all above 10 GeV, the strings for pp at 200 GeV reach much higher
masses. 

\begin{figure}
{\par\centering \resizebox*{1\columnwidth}{!}{\includegraphics{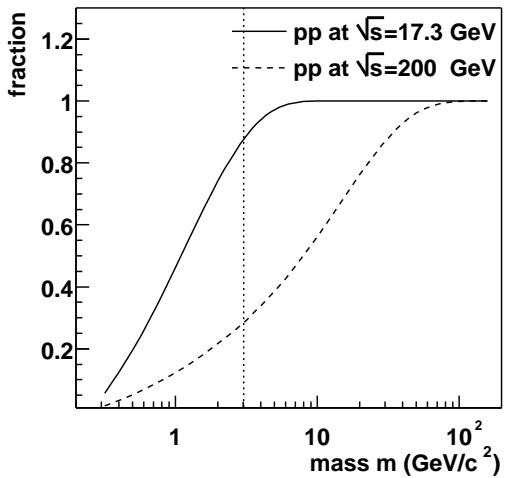}} \par}

\caption{\label{fig:5}Integrated string masses: Shown is the relative fraction of strings
below a certain mass \protect\( m\protect \). The line at 3 GeV shows the threshold
for \protect\( \Omega \protect \) production.}
\end{figure}
More conclusive is figure \ref{fig:5} where we see the corresponding cumulative
distributions, i.e. the fraction 
\[
F(m)=\frac{\int _{0}^{m}\frac{dn}{dm'}dm'}{\int _{0}^{\infty }\frac{dn}{dm'}dm'}\]
of strings with masses below \( m \). At 17.3 GeV 50\% of strings are lighter
than 1 GeV, at 200 GeV the fraction is only 10\%. Strings below 1 GeV cannot
produce any baryons. If we want to create a \( \Omega  \) given a \( \mathrm{s}-\bar{\mathrm{d}} \)
string, we will find in addition a \( \bar{\Xi } \), since we have to break
the string with the creation of a \( \mathrm{ss}-\bar{\mathrm{s}}\bar{\mathrm{s}} \)
pair. Therefore the minimum mass is above 3 GeV in the best case scenario, where
one strange quark is already given by the initial string. Consequently, it is
hard to create \( \Omega  \)'s at low beam energies since only 10\% of the
strings have the necessary mass, whereas at RHIC-energies 70\% of the strings
could kinematically produce \( \Omega  \)'s.

\section{Conclusions}

We can conclude that the different masses for the strings at the different beam
energies are responsible for a possible suppression of heavy hadrons in pp collisions
as compared to \( \mathrm{e}^{+}\mathrm{e}^{-} \)annihilation. If the hadron
mass is small as compared to the typical string energy the hadron multiplicity
ratios reach asymptotic values. A further increase of the string energy leads
only to an overall increase of the produced hadron multiplicity leaving their
relative ratio unchanged. If the string mass becomes comparable to the hadron
masses, the production of these hadrons are suppressed due to the very limited
phase space available.

This explains why in thermal fits \cite{becattini} a common constant temperature
has been found for high energy \( \mathrm{p}\bar{\mathrm{p}} \) collisions
and \( \mathrm{e}^{+}\mathrm{e}^{-} \)annihilation. The physical origin of
this phenomenon has in our model nothing to do with the formation of a thermal
system. Therefore it is premature to identify this fit parameter with a true
temperature. Employing a string fragmentation model which describes the kinematical
variables as well the multiplicities of particle species in \( \mathrm{e}^{+}\mathrm{e}^{-} \),
\( \mathrm{pp} \) and \( \mathrm{p}\bar{\mathrm{p}} \) collisions allows to
interpret the results of \cite{becattini} in physical terms without touching
the claim that data can be well fitted using the functional forms of a grand
canonical description.


\begin{thebibliography}{1}
\bibitem{rafelski}J. Rafelski, Phys.Rept.88:331,1982 , Phys. Rev. Lett. 48:1066 (1982)
\bibitem{WA97}F. Antinori et al., Nucl.Phys.A661:130 (1999)
\bibitem{nexus}H.J. Drescher et al., hep-ph/0007198 accepted by Phys. Rep., Phys.Rev.Lett.86:3506
(2001)
\bibitem{opal}Opal collaboration: R. Ackers et al., Zeit.Phys. C63:181 (1994), G. Alexander
et al., Zeit.Phys. C73:569 (1997), Zeit.Phys. C73:587 (1997)
\bibitem{pp19}M. Gazdzicki, Nucl.Phys. A528:754 (1991)
\bibitem{pp200}UA5 Collaboration, R.E. Ansorge et al., Nucl.Phys. B328:36 (1989) 
\bibitem{becattini}F. Beccattini, Z.Phys.C76:269 (1997), Z.Phys.C72:491 (1996) 
\end{thebibliography}
\end{document}